\documentclass{webofc}

\usepackage[varg]{txfonts}   % Web of Conferences font
\usepackage{hyperref}
\usepackage{url}
\usepackage{amssymb}
\usepackage{amsmath}
\hypersetup{colorlinks=true, linkcolor=blue, citecolor=blue, urlcolor=blue}

\hypersetup{colorlinks=true,citecolor=blue,urlcolor=blue,linkcolor=blue}
\usepackage{microtype}
\newcommand{\rmi}[1]{{\mbox{\scriptsize #1}}}
\newcommand{\rmii}[1]{{\mbox{\tiny\rm{#1}}}}

\def\gsim{\mathrel{\rlap{\lower0.25em\hbox{$\sim$}}\raise0.2em\hbox{$>$}}} 
\def\lsim{\mathrel{\rlap{\lower0.25em\hbox{$\sim$}}\raise0.2em\hbox{$<$}}}
\def\vec#1{\text{\boldmath$#1$}}
\def\nr#1{(\ref{#1})}
\def\be{\begin{equation}}
\def\ee{\end{equation}}
\def\bea{\begin{eqnarray}}
\def\eea{\end{eqnarray}}
\begin{document}
\title{
  Theory overview: electroweak emission from heavy-ion\\ collisions\footnote{
Invited plenary talk at the 31st International Conference on Ultra-relativistic Nucleus-Nucleus Collisions 
(Quark Matter 2025), Frankfurt, Germany, Apr 6-12, 2025.
  }
}

\author{\firstname{Greg} \lastname{Jackson}\inst{1}\fnsep\thanks{\email{jackson@subatech.in2p3.fr}} 
}

\institute{
SUBATECH 
  (IMT Atlantique, 
Nantes Universit\'e, 
  IN2P3/CNRS), \\ 
4 rue Alfred Kastler, 
La Chantrerie BP 20722, 
44307 Nantes, France 
          }

\abstract{
An important class of observables in 
the heavy-ion collision programme 
concerns %is provided by 
probes which are not sensitive to the prevailing strong interactions of QCD. 
The emission of photons, weak gauge bosons, and leptons fall into this 
category. 
Here I will describe the current status of such investigations, 
focusing on the emerging  
 theoretical picture and its uncertainties. 
}
\maketitle
\section{Introduction}
\label{intro}

For weak probes of strongly interacting systems, 
the hierarchy $\alpha_\rmi{em} \ll \alpha_s$
allows 
the calculation 
to be conveniently organised 
by first expanding the result in $\alpha_\rmi{em} = e^2/(4\pi)\,$.
For instance, photons are generated 
(via $H_\rmi{int} =  i\,e\, A_\mu J^{\,\mu}_\rmi{em}\,$)
at a rate given
by~\cite{Laine:2016hma,Kapusta:2006pm}
\be
  \Gamma_\gamma 
  \; \sim \; 
  e^2 \, \big\langle \, {{  J_\rmi{em} \,   J_\rmi{em}}} \, \big\rangle + {\cal O}(e^4)
  \, ,
  \label{eq:1}
\ee
where the average is calculated from the density matrix 
$\langle ... \rangle = {\rm Tr}\big[\, \hat \rho\, (...)\,\big]\,$. 
In heavy-ion collisions the system is too short-lived for 
photons or other weak probes to equilibrate, 
so they ``escape''
basically unmodified, 
unlike jets or soft hadrons 
(see refs.~\cite{Gale:2025ome} and \cite{Geurts:2022xmk} for 
recent overviews). 
Produced at all stages and throughout the volume, 
they are most copious when the medium is hottest, 
posing the challenge of disentangling the various sources.

The theoretical agenda is twofold. 
It involves $i$) the microscopic calculation of 
production rates, as well as $ii$) the macroscopic 
embedding 
of those rates in simulations of heavy-ion collisions.
Task $i$) requires field theory calculations of eq.~\nr{eq:1}, 
such as by expanding in $\alpha_s$, 
using effective hadronic or QCD based theories, 
or from non-perturbative lattice studies.
Predictions, e.g. for the $p_\rmii{T}$ spectrum of photons 
and the invariant mass $M$ distribution of dileptons, 
are obtained from $ii$) by integrating 
over the spacetime history. 
This second step must be 
done numerically 
using multi-stage frameworks, which 
typically involve
hydrodynamic codes such 
as {\small MUSIC}~\cite{Gale:2021emg} 
or {\em Trajectum}~\cite{Massen:2024pnj}.

In these proceedings, I will highlight recent theoretical progress 
on electroweak probes as presented at the conference.  
Let me draw attention to the excellent similar reports in 
refs.~\cite{Du:2025ith,Vujanovic:2024bby,Paquet:2023vkq,Tripolt:2020dac}, 
from the past few years. 
For an update of experimental developments, see ref.~\cite{Sebastian}. 
Here I will focus on photons and dileptons, 
using a notation where $\Gamma_\gamma$  
and $\Gamma_{\ell \bar \ell}$
are the respective 
differential rates per time and volume. 
The dilepton pair is produced through a virtual photon, 
of energy $\omega \geq ({ 4 m_\ell^2 + \vec{k}^2})^{1/2}$, 
where $\vec{k}$ is the photon's momentum 
and $m_\ell$ is the mass of each lepton. 
For real photons, $\omega = k \equiv | \vec{k} |\,$.

\section{Main stages and their emission rates}
\label{sec-1}

A source of electromagnetic probes which is common to both 
pp and AA collisions, 
originates from the first instant of the collision 
where 
eq.~\eqref{eq:1} 
involves the initial hadronic wavefunction of the projectiles. 
Such photons and dileptons are called {\em prompt} and 
can be calculated within the QCD factorization framework, 
involving the parton distribution functions (PDFs, $f_{i/\rmi{A}}$ below)~\cite{Arleo:2011gc}.
For dileptons this is called the Drell-Yan process, 
and at leading order (LO) it is
\footnote{
  Equation~\eqref{eq:dy} only includes the $\gamma^*$ intermediate state,
  but for large $M\,$, the $Z$ boson must also be taken into account. 
}
\be
  \frac{{\rm d}N^\rmi{DY}_{\rmi{AB} \to \ell \bar \ell}}{{\rm d}M^{\,2} \, {\rm d}y}
  \bigg|_\rmi{LO}
  \; =  \; 
  T_\rmi{AB}
  \, \frac{ e^4 
  \, {\textstyle{\sum^{ }_{i} }} 
  Q^2_{i} }{36 \pi \, M^{\,2} s}
  \ f^{ } _{i/\rmi{A}}(x_1)
  \,f^{ }_{\bar{i}/\rmi{B}}(x_2) 
  \; ,
  \label{eq:dy}
\ee
the geometric overlap $T_\rmi{AB} = N_{\rm coll}/ \sigma^\rmi{inel}_{\rm pp}$ 
is an effective luminosity, 
the quark charge fractions
are $Q_i = \big\{ \tfrac23, - \tfrac13 \big\}\,$, 
and nuclear PDFs 
(per nucleon)
are evaluated at 
the longitudinal momentum fractions 
$x_{1,2} = M e^{\pm y}/\sqrt{s}$ ($y$ is the rapidity of the dilepton).  
At higher order in perturbation theory, 
eq.~\eqref{eq:dy} generalises to a convolution of the nuclear PDFs 
with a hard scattering kernel and the kinematics is more complicated.\footnote{
  The formalism introduces a renormalisation scale $\mu_\rmii{R}$ (e.g. in $\alpha_s$) 
  and a factorization scale $\mu_\rmii{F}$ (e.g. in $f_{i/\rmi{A}}$) 
  which are typically evaluated at %the ``hard scale'' 
  $\mu_\rmii{R}\, , \mu_\rmii{F} = M\,$. 
  This choice is varied by a factor of $2$ to estimate the uncertainty.
} 
The Drell-Yan signal is dominant for large $M$ and 
allows one to study isospin content (via $Q_i$), 
sea quark asymmetry, 
and possibly the small-$x$ behaviour of nuclear PDFs (at large/small $y$).

\begin{figure}
\centering
\sidecaption
\includegraphics[width=6.5cm,clip]{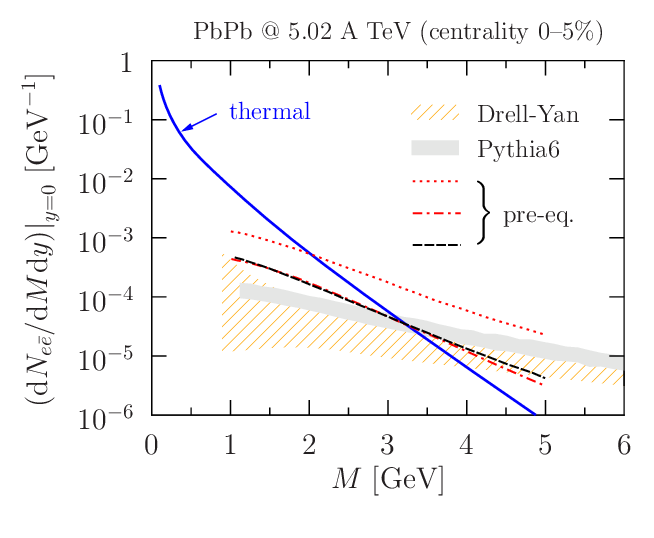}
\quad
\caption{
  Dielectron yield from the QGP as a function of $M\,$, 
  for central PbPb collisions %at the LHC 
  at midrapidity. 
  The thermal component (blue solid) 
  represent NLO+LPM$^{\rm LO}$
  convoluted with hydrodynamics 
  (see fig.~3 of ref.~\cite{Wu:2024pba}). 
  The pre-equilibrium 
  results include the LO rates in kinetic theory 
  with (red dash-dotted, black dashed) and 
  without (red dotted) initial quark suppression.
  See ref.~\cite{Garcia-Montero:2024lbl} (black, event-by- event simulation) and 
  ref.~\cite{Coquet:2021lca} (red, boost invariant evolution) for more details. 
  The NLO Drell-Yan result is shown, as well as a 
  prediction from the Pythia6 event generator.
  }
  \vspace{-.5cm}
\label{fig-1}
\end{figure}

Quite rapidly (within a time $\tau_\rmi{eq} \simeq 1$~fm/$c$), 
the longitudinal initial momentum distributions of partonic 
degrees of freedom  %are understood to 
approach local thermal equilibrium. 
Considerable progress in understanding how 
this occurs 
has been made with 
 kinetic transport theory. 
In this approach, 
Boltzmann equations describe the evolution of  
 quark and gluon distribution functions (denoted $f_{q,\bar q,g}$ below)
 and can similarly track %be augmented to track  
photon emission~\cite{Garcia-Montero:2023lrd}.  
%For example, 
The rate from $2\to 2$ processes is
\be
  \frac{{\rm d}\Gamma^{ }_\gamma}{{\rm d}^3 \vec{k} } 
  \bigg|_{2 \to 2}
  \ = \ 
  \frac{1}{4 k (2\pi)^3}
  {\textstyle{\sum\limits^{ }_{a,b,c} }}
  \int \! {\rm d} 
  \Omega_{2\to 2} \, {{f_a}}\,  {{f_b}} \, [ 1 \pm {{f_c}} ] \, |{\cal M}_{ab\to c\gamma}|^2
  \; ,
\ee
where at LO one has $q \bar q \to g \gamma$ and $ qg \to q\gamma$. 
Also included are $1+n\leftrightarrow 2+n$ processes,
as mandated by the Landau-Pomeranchuk-Migdal (LPM) effect.
The rate $q \bar q \to \gamma^*$ has been used to describe pre-equilibrium dilepton yields.\footnote{
  Such a LO dilepton rate is similar to eq.~\eqref{eq:dy}, with the nPDFs replaced with general 
  distribution functions.
}
Those yields 
have a particular scaling in time 
which involves  
the viscosity to entropy density ratio~\cite{Garcia-Montero:2024lbl}. 
At LHC energies, the initial wavefunction is dominated by gluons and it takes 
some time to populate the quark and antiquark distributions. 
This implies a suppression of very early electroweak emissions, 
which 
It has been suggested 
may influence %be revealed in 
 the dilepton invariant mass spectrum~\cite{Coquet:2021lca,Wu:2024pba}, 
see fig.~\ref{fig-1}. 

In thermal equilibrium, 
 eq.~\eqref{eq:1} is evaluated with a thermal density matrix 
 characterized by a temperature $T$ (and chemical potentials). 
The differential rates are proportional to the 
vector spectral function, 
given formally by 
analytic continuation of the Euclidean correlator:
\be
  \rho_{ }^{\mu\nu} (\omega, \vec{k})
   = 
  - {\rm Im} 
  \big[
    \textstyle\int_0^{1/T} \!\! {\rm d}\tau \; 
  e^{ i \, k_n \tau }
    G_{ }^{\mu\nu}( \tau , \vec{k})
    \big]_{ik_n \to \omega + i\,0^+}
  \ ;
  \quad
    G_{ }^{\mu\nu} %( \tau , \vec{k}) 
    \equiv
    \textstyle\int_{\vec{x}} 
  e^{ - i \vec{k}\cdot\vec{x}}
 \, \big\langle \,
    V_{ }^{\,\mu}(\tau , \vec{x})
    V_{ }^{\,\nu}(0)
 \, \big\rangle \,
  \label{eq:spec}
\ee
\if0
\be
  \rho^{\mu\nu} (\omega, \vec{k})
  \; = \; 
  - {\rm Im} 
  \bigg[
    \int_{\vec{x}} 
  \int_0^{1/T} \!\! {\rm d}\tau \; 
  e^{ - i \, (k_n \tau + \vec{x}\cdot\vec{k})}
 \, \Big\langle \,
    V_{ }^{\,\mu}(\tau , \vec{x})
    V_{ }^{\,\nu}(0)
 \, \Big\rangle \,
    \bigg]_{ik_n \to \omega + i\,0^+}
  \, ,
  \label{eq:spec}
\ee
\fi
where $\int_{\vec{x}} \equiv \int {\rm d}^3 \vec{x}$ and
$V^{\,\mu} \equiv \bar \psi \gamma^\mu \psi\,$. 
The photon and dilepton rates are given, respectively by
\bea
   \frac{{\rm d}\Gamma_\gamma}{
    {\rm d}^3 {\vec{k}}}
  & = &
  \frac{\alpha_\rmi{em} \sum_i Q^{\,2}_{i}}{2\pi^2 \, k }
  n_\rmi{B}(k\,) \, 
      {{ \rho^{{}}_\rmi{V}\,(\omega, \vec{k})\big|_{\omega = k}  }}
  \, + \, {\cal O}(\alpha_\rmi{em}^2) 
  \, , 
  \label{eq:photon}
%\eea
%\bea
  \\
  \frac{{\rm d}\Gamma_{\ell \bar \ell}}{
    {\rm d}\omega {\rm d}^3 {\vec{k}}}
  & \approx &
  \frac{\alpha_\rmi{em}^{\,2} \sum_i Q^{\,2}_{i} }{3\pi^{\,3} M^{\,2}}
  n_\rmi{B}(\omega) \,
      {{ \rho^{{ }}_\rmi{V}\,(\omega, \vec{k}) }}
  \, + \, {\cal O}(\alpha_\rmi{em}^3) 
  \, , 
  \hspace{.8cm}
  \text{where}
  \quad 
  \rho_\rmi{V} \equiv \rho_{\mu}^{\ \,\mu}
  \; .
  \label{eq:dilepton}
\eea
Here $\omega$ and $\vec{k}$ 
are the photon energy and momentum in the 
medium's local rest frame.
However, eqs.~\eqref{eq:photon} and \eqref{eq:dilepton} 
apply cell-by-cell when simulating a heavy ion collision. 
The temperature $T(t,\vec{x})$ varies in space and time, while 
$\omega$ and $\vec{k}$ need to be boosted to the lab frame 
according to the local fluid velocity $u^{\,\mu}(t,\vec{x})\,$. 
Hydrodynamical evolution governs these macroscopic quantities, 
and hence the observed yields 
${\rm d}N = \int_{t,\vec{x}} {\rm d}\Gamma$ 
are sensitive to viscous corrections (as are the rates themselves)~\cite{Shen:2014nfa,Vujanovic:2017psb,Vujanovic:2019yih}. 
New results for finite $\mu_\rmi{B}$ were presented at the 
conference~\cite{Xian-Yu}.

As the system cools and expands, 
partons confine into a gas of hadrons. 
Particles are generated on a freeze-out surface, 
ending the thermal phase, 
and undergo hadronic re-scatterings 
and resonance decays. 
It is speculated that the hadronization process 
itself may entail photon production~\cite{Fujii:2022hxa}. 
The subsequent hadronic dynamics is usually 
treated with transport codes (e.g. SMASH or UrQMD).
Photons are produced by individual scatterings 
in these approaches, requiring %which require those 
 cross sections as input~\cite{Gotz:2021dco}. 
For lower collisional energies (e.g. the RHIC beam energy scan), 
 the bayon-rich %finite-$\mu_\rmi{B}$ 
region of the phase diagram is explored.
A ``cocktail'' of hadronic reactions show up in the 
 dilepton invariant mass spectrum, see e.g. ref.~\cite{Jorge:2025wwp}. 
Other novel phenomena may also occur as $\mu_\rmi{B}$ increases, 
for example a phase transition (rather than crossover)~\cite{Savchuk:2022aev} 
or a possible ``moat'' regime~\cite{Nussinov:2024erh}
could  lead to enhanced low mass dileptons.

\section{Perturbation theory and lattice}
\label{sec-2}

The spectral function in 
  eqs.~\eqref{eq:photon} and \eqref{eq:dilepton} 
  may be evaluated from first principles 
  either by a weak coupling expansion 
  (in $\alpha_s$) 
  or by non-perturbative lattice QCD.
Much of the phenomenology in sec.~\ref{sec-1} is currently based on 
perturbation theory.\footnote{
  The fixed value $\alpha_s \approx 0.3$ ($g \approx 2$) is commonly used,  
  although not always stated in the literature.
} 
In equilibrium, the nature of the 
strict perturbative calculation 
of $\rho_\rmi{V}$ (which has a long history) 
depends upon the 
kinematic domain. 
To compare with lattice data, 
the strict NLO result (applicable for $M \gsim T$) 
is combined with 
the regime of LPM resumation (for $M \sim gT$), 
 see ref.~\cite{Jackson:2019yao}. 
If $\rho^{\mu\nu}$ is known, 
\be
  G^{\mu\nu}_{ }(\tau,\vec{k})
  \; = \;
    {\textstyle{\int}}_0^\infty 
    %\tfrac{d\,\omega}{\pi}
    {\rm d} \omega \, 
    \rho^{\,\mu\nu}_{ } \big(\omega, \vec{k}\big)
    \, 
  %\frac{
    \cosh[(\tfrac1{2T}  - \tau)\omega]
  %}{
  \big/
  \big( \pi 
    \sinh [\tfrac{\omega}{2T}  ] \big)
  %} 
  \; .
\ee
The reverse, i.e. 
extracting the frequency dependence of $\rho$ from measurements 
of $G\,$, 
is an ill-conditioned operation for which there 
are a variety of strategies~\cite{Francis:2025elc}.
Studies were first made at
 zero momentum~($\vec{k}=0$)~\cite{Ding:2010ga} 
and then non-zero momentum~\cite{Ghiglieri:2016tvj}. 
Recent progress 
enables us to scrutinize pQCD and make quantitative estimates of  
non-perturbative contributions.\footnote{
  A different approach, free of the inverse problem, 
  utilises imaginary spatial momenta~\cite{Meyer:2018xpt,Ce:2023oak}.
}

The tensor structure has two independent components, 
transverse (T) and longitudinal (L) w.r.t. $\vec{k}\,$.
Aligning the momentum with the $z$-axis: $\vec{k} \to k\, \vec{e}_z\,$, 
they can be expressed as 
$
G_\rmi{T}^{ }
\equiv
\tfrac12 
 {\textstyle\sum_{i=1}^{2}} G_{ }^{ii}
 %( G_{ }^{11} + G_{ }^{22} )
$ 
and 
$
G_\rmi{L}^{ }
\equiv
 G_{ }^{33}
 -
 G_{ }^{00}
\,$.
Another basis %of correlators
is 
$%\be
 G_\rmi{V}^{ } 
 \; \equiv \;
 2\, G_\rmi{T}^{ } 
 +
 G_\rmi{L}^{ } 
$ and 
% \  , 
% \quad
 $G_\rmi{H}^{ } 
 \; \equiv \;
 2\, \big( G_\rmi{T}^{ } 
 -
 G_\rmi{L}^{ } 
 \big)
 %\; 
$. %\ee
For an $N_s^3 \times N_\tau$ lattice, 
the 
momenta
%correlators can be measured at 
$k = 2\pi n /(a N_s)\,$, where $a$ is the lattice spacing 
and $n$ is an integer. 
The correlators are periodic in the imaginary-time direction and
symmetric on the interval 
$\tau \in [0,1/T]\,$, 
although  cut-off effects are difficult to control   
for $\tau T \lsim \frac{1}{4}\,$.

\begin{figure}
\centering
\sidecaption
\includegraphics[width=6.5cm,clip]{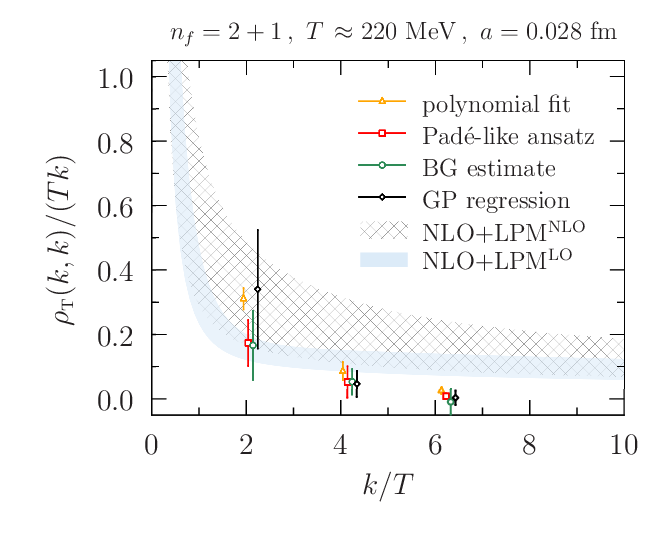}
\quad
\caption{
  Transverse spectral function 
  at the photon point $\omega = |\vec {k}|\,$, 
  obtained from spectral reconstruction 
  methods for $G_\rmi{H}$~\cite{Ali:2024xae}. 
  Results are shown for simulations 
  on a fixed lattice for $T = 220$~MeV 
  (spacing $a = 0.028$~fm), 
  with an unphysical
  pion mass of $m_\pi = 320$~MeV. 
  The correlators were 
  measured at $k = T \times 2\pi n/3\,$, $n=1,2,3\,$. 
  (Error bars are slightly displaced for better visibility;
  BG = Bakus-Gilbert; GP = Gaussian Process.) 
  The resummed spectral function from perturbation theory 
  is also shown as a band (which results from the scale variation in $\alpha_s$) 
  %$\mu = [\frac12 , 2]\mu_\rmi{opt}$), 
  see ref.~\cite{Jackson:2019yao} for details.
  }
\vspace{-.5cm}
\label{fig-2}
\end{figure}

Note that for $\omega = k\,$, the Ward identity 
implies $\rho^{00} = \rho^{33}$ and hence $\rho_\rmi{L} = 0\,$. 
Therefore either $G_\rmi{V}$ or $G_\rmi{H}$ may be analyzed for the photon rate. 
Dileptons however require $\rho_\rmi{V}$ in the timelike domain, 
where it receives a large contribution $\propto M^2 = \omega^2 - k^2$ 
from vacuum physics. 
This reflects itself in $G_\rmi{V} \sim 1/\tau^3$ for 
small $\tau\,$, making it difficult to constrain the {\em thermal} 
information in $\rho_\rmi{V}\,$. 
On the other hand, 
$\rho_\rmi{H}$ vanishes exactly in vacuum (as well as for $\vec{k} = 0$) and
satisfies the sum rule $\int {\rm d}\omega \, \omega \rho_\rmi{H} = 0\,$.  
There is now a wealth of information on Euclidean correlators and 
spectral functions %for $T > T_c$
for $T > T_c$~\cite{Ce:2020tmx,Ce:2022fot,Ali:2024xae}. 
Phenomenological models should therefore be tested on 
these lattice data before making predictions.
So far these studies seem to
indicate, see fig.~\ref{fig-2}, 
that pQCD alone slightly overestimates the 
photon rate at large $k/T$.

\section{What we can learn from EM probes}
\label{sec-3}

\subsection{Temperature of the QGP}

The thermal distribution function in 
eqs.~\eqref{eq:photon} and \eqref{eq:dilepton} 
 implies that  
the integrated rates scale like $T^4$, 
i.e. hotter regions radiate the most.  
For this reason photons and dileptons can 
act as ``thermometers'' of the medium. 
If the measured spectra are approximately exponential, 
an effective temperature $T_\rmi{eff}$ can be 
defined from the inverse logarithmic slope, e.g.
$%\be
  %\frac{{\rm d}N_\gamma}{{\rm d}y \, p_\rmii{T} {\rm d}p_\rmii{T}}
  {{\rm d}N_\gamma} / ({{\rm d}y \, p_\rmii{T} {\rm d}p_\rmii{T}})
  \; \propto \; 
  \exp (
    - 
    {p_\rmii{T}}/{T_\rmi{eff}}
    )
$
%  \; ,
%  \quad
 and 
$
  %\frac{{\rm d}N_{\ell \bar \ell}}{{\rm d}y \, {\rm d}M}
  {{\rm d}N_{\ell \bar \ell}}/({{\rm d}y \, {\rm d}M})
  \; \propto \; 
  (MT_\rmi{eff})^{3/2} \, 
  \exp (
    -  M / T_\rmi{eff}
    )
\,$.
  %\; .
%\ee
However, as clear from sec.~\ref{sec-1}, 
 the value of $T_\rmi{eff}$ derives
from a superposition of sources. 
It is important to understand which factors control
this value, in order to interpret the 
experimental findings~\cite{Sebastian}. 
In the case of photons, 
$p_\rmii{T}$ 
is blue shifted relative to its 
value in the local rest frame due to radial flow~\cite{Du:2024pbd,Massen:2024pnj}. 
For dileptons $M$ is an invariant quantity 
and the 
slope parameter 
is found to correlate with the 
initial mean temperature in simulations~\cite{Churchill:2023zkk,Churchill:2023vpt,Renan}.

\subsection{Electrical conductivity}

The limit $\omega, k \ll T$ 
is one in which charge fluctuations 
  are classical in nature 
and hence the spectral function takes the ``universal'' form: 
$%\be
  \rho_\rmi{V}(\omega,\vec{k}\,)
  \; = \; 
  \big(
  \tfrac{\omega^2 - k^{\,2}}{\omega^2 + {D}^{\,2} k^{\,4}}
  + 2
  \big) 
  \ \chi_\rmi{q}  
  \, D
  \, \omega
  %\sigma_\rmi{el} 
  %\; ,
  %\label{eq:hydro}
$, %\ee
where $D$ is the electrical diffusion coefficient and 
$
  \chi_\rmi{q} 
  = 
  \frac{1}{T}
  G^{00}(\tau,\vec{0})
  %= \, \frac{1}{T}\,\int_{\vec{x}} \big\langle 
  %V^{\,0}(\tau,\vec{x}) V^{\,0}(0)
  %\big\rangle
  %\! = \; \chi_\rmi{q} T
  $
is the quark susceptibility. 
The classical (hydrodynamic) description 
involves parameters %, called transport coefficients, 
which need to be matched to the microscopic theory, 
e.g. 
the electrical conductivity $\sigma_\rmi{el}\,$ defined by 
$\langle \vec{J}_\rmi{em} \rangle = \sigma_\rmi{el} \vec{E}\,$.
One can obtain the value of the latter from 
certain limits, % of eq.~\eqref{eq:hydro}, 
as given by the 
Kubo formulae
\be
  \sigma_\rmi{el} 
  \ = \ 
  \chi_\rmi{q} \, D
  \ = \ 
  \tfrac12 \,
  %\lim_{k \to 0}
  {\textstyle{\lim\limits^{ }_{k\to 0} }} \ 
  \rho_\rmi{V} (k, \vec{k}\,)\,\big/\,k
  \ = \ 
  \tfrac13 \,
  %\lim_{\omega \to 0}
  {\textstyle{\lim\limits^{ }_{\omega\to 0} }} \ 
  \rho_\rmi{V} (\omega, \vec{0})\,\big/\,\omega
  \; .
\ee
For this reason, is has been suggested that 
$\sigma_\rmi{el}$ can be obtained from the appropriate limit 
of photon and dilepton spectra~\cite{Floerchinger:2021xhb}. 
For dileptons, the hadronic component is paramount in this limit 
and in-medium modifications to vector mesons 
need to be treated accurately~\cite{Rapp:2024grb,Atchison:2024lmf}.

\subsection{Dilepton polarization}

\begin{figure}
\centering
\sidecaption
\includegraphics[width=6.5cm,clip]{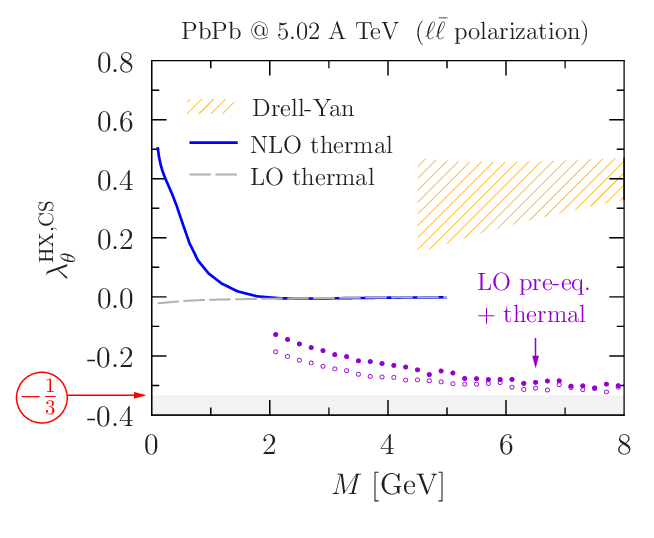}
\quad
\caption{
  The angular coefficient $\lambda_\theta$ for dileptons in 
  different frames, 
  as a function of invariant mass.
  Thermal QGP results are from ref.~\cite{Wu:2024vyc}~(0--20\% centrality), 
  and computed in the helicity frame. 
  The pre-equilibrium and Drell-Yan curves 
  are computed in the Collins-Soper frame, see
  ref.~\cite{Coquet:2023wjk}~(0--5\% centrality). 
  Both the LO thermal and pre-equilibrium 
  originate from $q \bar q \to \gamma^*$. 
  (The open circles have a longer equilibration 
  time than the filled ones.)
  }
\vspace{-.7cm}
\label{fig-3}
\end{figure}

The largest theoretical uncertainty in the succession of models 
described in sec.~\ref{sec-1} is 
the earliest timespan where the system 
is highly anisotropic. 
This far from equilibrium state 
is difficult to treat rigorously, 
but impacts the polarization of photons~\cite{Hauksson:2023dwh}. 
The angular distribution of dileptons 
 has drawn attention recently, being sensitive to 
polarization~\cite{Speranza:2018osi}, 
and is usually parametrized 
(by the leading, frame dependent, coefficients  
$\lambda_\theta$ and $\lambda_\phi$)
by 
\be
   \frac{{\rm d}N_{\ell \bar \ell}}{
     {\rm d} M \, {\rm d} y \, 
     %{\rm d}\omega \, {\rm d}^3 \vec{k} \, 
     %{\rm d}\cos \theta_\ell \, {\rm d} \phi_\ell
   {\rm d}\Omega_\ell 
 } 
   \ \propto \ 
   1 
   + {{\lambda_\theta}} \cos^2 \theta_\ell 
   + {{\lambda_\phi}} \sin^2\theta_\ell \cos 2\phi_\ell 
   + \ldots 
   \; , 
   \label{eq:pol}
\ee
where the angles ${\rm d}\Omega_\ell  = {\rm d}\cos \theta_\ell \, {\rm d} \phi_\ell$ 
refer to one of the final leptons, 
measured w.r.t. a $z$-axis chosen conventionally in the virtual photon's rest frame. 

In the helicity frame (HX), 
$\lambda^\rmi{HX}_\theta 
\simeq
  (\rho_\rmi{T} - \rho_\rmi{L}
  )/( 
  \rho_\rmi{T} + \rho_\rmi{L})\,$ 
which has been studied in both hadronic~\cite{Seck:2023oyt} 
and QGP~\cite{Wu:2024vyc} phases. 
The coefficient $\lambda^\rmi{HX}_\theta$ 
receives a large NLO correction for $M \lsim 2$~GeV~(fig.~\ref{fig-3}). 
In the Collins-Soper (CS) frame, 
$\lambda^\rmi{CS}_\theta = 3 Q/ (\frac25 - Q)$ 
where $Q$ is the 
quadrupole moment 
of the lepton distribution, 
which was calculated 
for the LO pre-equilibrium 
process $q \bar q \to \ell \bar \ell$~\cite{Coquet:2023wjk}. 
When the initial quarks 
have large longitudinal momenta: 
$\lambda^\rmi{CS}_\theta = 1\,$%(for Drell-Yan at LO). 
At the other extreme, when the initial quarks 
carry purely transverse momentum: 
$\lambda^\rmi{CS}_\theta = -\frac13\,$. 
A sign change of $\lambda_\theta^\rmi{CS}$  
would thus distinguish these scenarios, see fig.~\ref{fig-3}. 

\vspace{-1mm}

\section{Summary}

Electroweak probes are valuable tools 
for diagnosing the medium in heavy ion collisions. 
Perturbative and lattice approaches 
now complement each other, 
enabling quantitative estimates of non-perturbative effects in emission rates. 
New insights are being provided by differential observables, like dilepton polarization.
Yet some puzzles remain, for example:
Measurements of the $v_2$ for direct photons suggests 
that current predictions are underestimating 
thermal photons. 
This is in tension with fig.~\ref{fig-2}, 
where lattice analyses indicate that 
perturbation theory is overestimating thermal photons.
Another questions is the scaling 
of direct photons with $({\rm d} N_\rmi{ch}/{\rm d}y)^\alpha$, 
although there is some disagreement about the 
value of $\alpha$ on the experimental side~\cite{Sebastian}. 
Diverse as the theoretical progress has been, 
there is still room (and even need) for 
contributions 
on both formal aspects and phenomenology. 

%----------
\vspace{0.5\baselineskip}
\begin{acknowledgement}
\footnotesize
I am very grateful to  
F.~Arleo, 
D.~Bala, 
M.~Coquet, 
L.~Du, 
C.~Gale, 
J.~Ghiglieri 
and
M.~Laine 
for their 
valuable 
comments. %on this report.
The author is supported by the Agence Nationale de la Recherche, 
under grant ANR-22-CE31-0018. %(AUTOTHERM). 
\end{acknowledgement}

\end{document}